\documentclass[twocolumn,showpacs,preprintnumbers,prb]{revtex4}
\usepackage{amssymb}
\usepackage{amsmath}
\usepackage{graphicx}

\begin{document}

\title{Crossover from Majorana edge to end states in quasi-one-dimensional $%
p $-wave superconductors}
\author{Bin Zhou$^{1,2}$, and Shun-Qing Shen$^{2}$}
\affiliation{$^{1}$Department of Physics, Hubei University, Wuhan 430062, China\\
$^{2}$Department of Physics and Center for Theoretical and Computational
Physics, The University of Hong Kong, Pokfulam Road, Hong Kong, China}
\date{\today}

\begin{abstract}
In a recent work [Potter and Lee, Phys. Rev. Lett. \textbf{105}, 227003
(2010)], it was demonstrated by means of numerical diagonalization that the
Majorana end states can be localized at opposite ends of a sample of an
ideal spinless $p$-wave superconductor with the strip geometry beyond the
strict one-dimensional limit. Here we reexamine this issue, and study the
topological quantum phase transition in the same system. We give the phase
diagrams of the presence of Majorana end modes by using of $Z_{2}$
topological index. It is found that the topological property of a strip
geometry will change in an oscillatory way with respect of the sample width.
\end{abstract}

\maketitle

\section{Introduction}

Recently, there is a growing interest in searching for Majorana fermions in
the condensed matter physics community.\cite{WilczekNP09} Unlike traditional
fermions, Majorana fermions are their own antiparticles and are expected to
obey non-Abelian statistics.\cite{ReadPRB00,IvanovPRL01} Due to novel
properties of the Majorana particles, it has been proposed that they can be
used in realizing the fault-tolerant topological quantum computation
protected from local decoherence.\cite%
{NayakRMP08,KitaevPU01,SternPRB04,AkhmerovPRB10,HasslerNJP10} The early
experimental efforts were focused on the $\gamma =5/2$ fractional quantum
Hall state,\cite{MooreNPB91,GreiterNPB92} in which the Majorana fermions are
expected to be charge $e/4$ quasiholes. Other proposed candidate materials
include the $p$-wave superconductors,\cite%
{ReadPRB00,SarmaPRB06,RoyPRL10,VolovikJETPL99} in which the Majorana
fermions are zero energy single-particle states trapped in vortex cores, and
superfluids in the $^{3}$He-$B$ phase.\cite{VolovikBook03,SilaevJETP10}
Recent several new suggestions are proposed to look for Majorana fermions,
such as topological insulators proximate to an $s$-wave superconductor,\cite%
{FuPRL08,FuPRB09,AkhmerovPRL09,TanakaPRL09,LawPRL09,LinderPRL10,ShivamoggiPRB10,IoselevichPRL11,NilssonPRL08}
a semiconducting thin film sandwiched between an $s$-wave superconductor and
a magnetic insulator,\cite{SauPRL10,LinderPRB10,AliceaPRB10} one-dimensional
(1D) semiconductor-superconductor heterostructures based on quantum wires,%
\cite{LutchynPRL10,OregPRL10,AliceaNP11} noncentrosymmetric superconductors
with broken time-reversal symmetry,\cite{FujimotoPRB08,GhoshPRB10} and $s$%
-wave superfluid of ultracold atoms.\cite{ZhuPRL11} Although extensive
efforts have been made, direct evidence of Majorana fermions is still
absent. Thus, it is still a challenging mission to produce and detect the
mysterious Majorana particles.

In his pioneering work, Kitaev\cite{KitaevPU01} found that Majorana
particles can be localized at the ends of a 1D $p_{x}+ip_{y}$
superconducting wire. In a recent paper,\cite{PotterPRL10} Potter and Lee
moved beyond the strict 1D limit and explored further the notion of Majorana
end states in quasi-1D $p_{x}+ip_{y}$ superconductors. They considered a
spinless $p$-wave superconductor with finite width, and showed by means of
numerical diagonalization that the Majorana end states can be localized at
opposite ends of a sample with the strip geometry. We also note that a
similar set of ideas on Majorana fermions in the quasi-1D wires was dicussed
by Wimmer et al\cite{WimmerPRL10} (to see Appendix C of the supplement of
Ref. 35). Very recently, searching for Majorana fermions in multiband
semiconducting nanowires attracts much attention and also opens a new path
in the challenging field.\cite%
{LutchynPRL11,PotterPRB11,LawXXX11,LutchynXXX11} In this paper, motivated by
the work by Potter and Lee,\cite{PotterPRL10} we reexamine an ideal spinless
quasi-1D $p_{x}+ip_{y}$ superconductor, and study the topological quantum
phase transition in the this system. Firstly, based on the tight-banding
model, we give Majorana edge modes localized at boundaries of a sample with
the sufficiently large width, and through analytical solutions in a model
calculation for a strip of finite width, we find that two Majorana fermions
on the two edges can couple together to produce a gap in the excitation
spectrum under the periodic boundary condition along the longitudinal
direction. Then, we give the phase diagrams of the presence of Majorana end
modes in quasi-1D sample by using of $Z_{2}$ topological index. We find that
the topological property of a strip geometry will change in an oscillatory
way with respect of the sample width.

The paper is organized as follows: in Sec. II, we shall give a
square-lattice tight-binding model to describe a spinless $p$-wave
superconductor. In Sec. III, we shall investigate the behaviors of Majorana
edge states localized at boundaries of the sample with finite width, and the
energy gap opened due to the coupling of two Majorana fermions at opposite
boundaries is given through analytical solutions in a model calculation.\cite%
{Note} In Sec. IV, we shall calculate the $Z_{2}$ topological index and plot
the phase diagrams of the presence of Majorana end modes in the quasi-1D $p$%
-wave superconductors. By analyzing the phase diagrams, we shall then give
the findings in the topological quantum phase transition in quasi-1D
spinless $p$-wave superconductors based on the tight-binding model. We shall
summarize our conclusions in Sec. V.

\section{Model}

We consider a two-dimensional (2D) Kitaev model of spinless $p$-wave
superconductors on a square-lattice, which is described by the following
Hamiltonian\cite{KitaevPU01,PotterPRL10}
\begin{eqnarray}
H &=&\sum_{j=1}^{L}\sum_{\alpha =1}^{n}\left[ -\mu c_{j,\alpha }^{\dag
}c_{j,\alpha }-\left( tc_{j,\alpha }^{\dag }c_{j,\alpha +1}+\left\vert
\Delta \right\vert c_{j,\alpha }c_{j,\alpha +1}\right. \right.  \notag \\
&&\left. \left. +tc_{j,\alpha }^{\dag }c_{j+1,\alpha }+i\left\vert \Delta
\right\vert c_{j,\alpha }c_{j+1,\alpha }+h.c.\right) \right] ,  \label{H1}
\end{eqnarray}%
where $c_{j,\alpha }^{\dag }$ creates an electron on site $(j,\alpha )$, $t$
($>0$) is the hopping amplitude, $\mu $ is the chemical potential, $%
\left\vert \Delta \right\vert $ ($>0$) is the $p$-wave pairing amplitude,
and the lattice spacing is taken as unity. Here we assume a strip geometry
in which the lattice site numbers are $L$ along the $x$-axis direction and $%
n $ along the $y$-axis direction (the sample width direction), thus which
totals $N=nL$ fermionic sites. First, one introduces a periodic boundary
condition along the $x$-axis direction, i.e., $c_{L+1,\alpha }^{\dag
}=c_{1,\alpha }^{\dag }$, and uses the Fourier transform of the operator $%
c_{j,\alpha }^{\dag }$%
\begin{equation}
c_{j,\alpha }^{\dag }=\frac{1}{\sqrt{L}}\sum_{q}c_{\alpha }^{\dag }\left(
q\right) e^{-iqj},
\end{equation}%
where $q=q_{x}$ is the wave vector along the $x$-axis, and $-\pi \leq q\leq
\pi $. In terms of the new creation and annihilation operators $c_{\alpha
}^{\dag }\left( q\right) $ and $c_{\alpha }\left( q\right) $, the
Hamiltonian (\ref{H1}) can be rewritten as \
\begin{eqnarray}
H &=&\sum_{q}\sum_{\alpha =1}^{n}\left\{ -\left( \mu +2t\cos q\right)
c_{\alpha }^{\dag }\left( q\right) c_{\alpha }\left( q\right) \right.  \notag
\\
&&-\left[ tc_{\alpha }^{\dag }\left( q\right) c_{\alpha +1}\left( q\right)
+\left\vert \Delta \right\vert c_{\alpha }\left( q\right) c_{\alpha
+1}\left( -q\right) \right.  \notag \\
&&\left. \left. +i\left\vert \Delta \right\vert e^{-iq}c_{\alpha }\left(
q\right) c_{\alpha }\left( -q\right) +h.c.\right] \right\} .  \label{H2}
\end{eqnarray}%
Then, we define a set of the operators $\gamma _{2\alpha -1}\left( q\right) $
and $\gamma _{2\alpha }\left( q\right) $ as
\begin{equation}
\gamma _{2\alpha -1}\left( q\right) =i\left[ c_{\alpha }^{\dag }\left(
-q\right) -c_{\alpha }\left( q\right) \right] ,
\end{equation}%
\begin{equation}
\gamma _{2\alpha }\left( q\right) =c_{\alpha }^{\dag }\left( -q\right)
+c_{\alpha }\left( q\right) ,
\end{equation}%
which satisfies the anticommutation relation $\left\{ \gamma _{m}^{\dag
}\left( q\right) ,\gamma _{n}\left( q^{\prime }\right) \right\} =2\delta
_{mn}\delta _{qq^{\prime }}$ and $\gamma _{m}^{\dag }\left( q\right) =\gamma
_{m}\left( -q\right) $. In fact, $\gamma _{m}\left( 0\right) $ is just a
Majorana fermion operator due to $\gamma _{m}^{\dag }\left( 0\right) =\gamma
_{m}\left( 0\right) $. In the basis of the news operators $\gamma _{2\alpha
-1}\left( q\right) $ and $\gamma _{2\alpha }\left( q\right) $, the
Hamiltonian (\ref{H2}) has the following form
\begin{equation}
H=i\frac{1}{4}\sum_{q}\sum_{\eta ,\kappa }B_{\eta ,\kappa }\left( q\right)
\gamma _{\eta }\left( -q\right) \gamma _{\kappa }\left( q\right) ,
\label{H3}
\end{equation}%
where the elements of the $2n\times 2n$ matrix $B\left( q\right) $ are given
as%
\begin{equation}
B_{2\alpha ,2\alpha }=-B_{2\alpha -1,2\alpha -1}=-2i\left\vert \Delta
\right\vert \sin q,
\end{equation}%
\begin{equation}
B_{2\alpha ,2\alpha -1}=-B_{2\alpha -1,2\alpha }=-\mu -2t\cos q,
\end{equation}%
\begin{equation}
B_{2\alpha ,2\alpha +1}=-B_{2\alpha +1,2\alpha }=-t-\left\vert \Delta
\right\vert ,
\end{equation}%
\begin{equation}
B_{2\alpha -1,2\alpha +2}=-B_{2\alpha +2,2\alpha -1}=t-\left\vert \Delta
\right\vert ,
\end{equation}%
and the else elements are zero.

\begin{figure}[tbp]
\includegraphics[width=8cm]{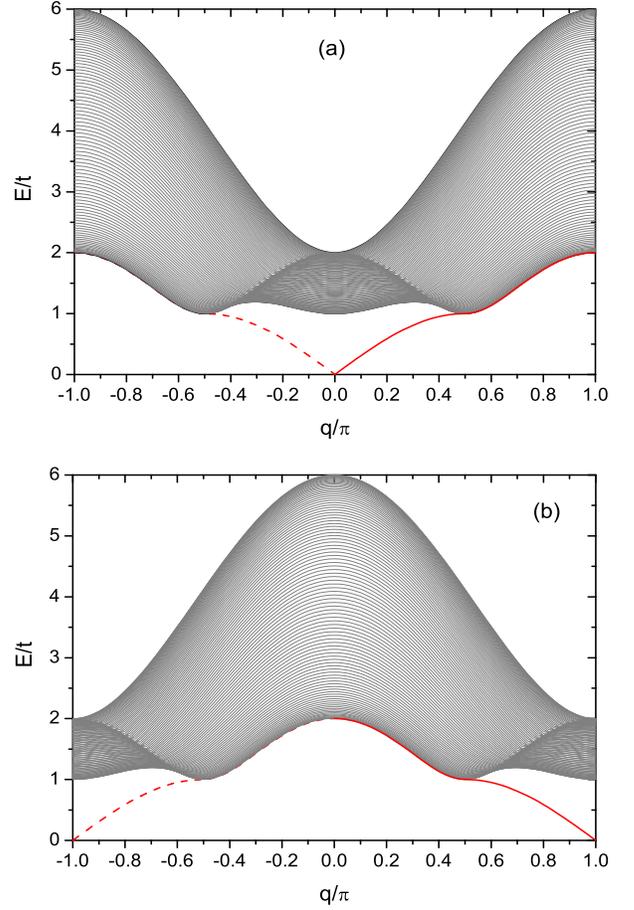}
\caption{${}$(color online). The excitation spectrum obtained by numerical
diagonalization of the Hamiltonian (\protect\ref{H2}) with the parameters $%
n=100$, $\Delta /t=0.5$. (a) $\protect\mu /t=-2.0$; (b) $\protect\mu /t=2.0$%
. The (red) solid and dashed lines correspond to the edge modes, which are
localized at opposite boundaries of the sample, respectively.}
\end{figure}

If we consider also the periodic boundary condition along the $y$-axis
direction, the bulk excitation spectrum of a 2D square-lattice Kitaev model
is given by
\begin{eqnarray}
E\left( q\right) &=&\left\{ \left[ 2t\left( \cos q_{x}+\cos q_{y}\right)
+\mu \right] ^{2}\right.  \notag \\
&&\left. +4\left\vert \Delta \right\vert ^{2}\left( \sin ^{2}q_{x}+\sin
^{2}q_{y}\right) \right\} ^{1/2}.  \label{SP}
\end{eqnarray}%
In general the spectrum (\ref{SP}) has a gap, but the gap will close when $%
\mu =-4t$ at $\mathbf{q=}\left( q_{x},q_{y}\right) =\left( 0,0\right) $ or $%
\mu =4t$ at $\mathbf{q}=\left( \pm \pi ,\pm \pi \right) $ or $\mu =0$ at $%
\mathbf{q}=\left( 0,\pm \pi \right) $ and $\mathbf{q}=\left( \pm \pi
,0\right) $. Actually, $\left\vert \mu \right\vert =4t$ is a phase
transition line. For a strip geometry with the periodic boundary condition
along the $x$-axis direction and the open boundary condition along the $y$%
-axis direction, if the sample width is sufficiently large (i.e., the
lattice site numbers $n$ is much larger than superconducting coherence
length $\xi _{0}\sim t/\left\vert \Delta \right\vert $), a pair of gapless
chiral edge modes per boundary always present for $-4t<\mu <0$ and $0<\mu
<4t $. In the former parameter region, the gapless points are present at $%
q=q_{x}=0$, and in the latter at $q=\pm \pi $. The excitation spectra
obtained by numerical diagonalization of the Hamiltonian (\ref{H2}) with $%
n=100$ are plotted in Fig. 1, where the chiral edge modes are shown inside
the bulk gap and the energies of chiral edge modes are $\pm 2\left\vert
\Delta \right\vert \sin q$. It is interesting to note that different from
the continuum model the minimum of the edge dispersion shifts from $q=0$ to $%
q=\pi $ as one goes from negative to positive $\mu $ in the tight-binding
model. This result can be understood based on the fact that if in the whole
1D Brillouin zone ($-\pi \leq q\leq \pi $) the gapless points are present at
both $q=0$ and $\pm \pi $, the edge modes with the spectrum $\pm 2\left\vert
\Delta \right\vert \sin q$ would intersect Fermi energy (assumed inside the
excitation gap) four times (an even multiple of two) which violates the
topological stability.\cite{WuPRL06,NielsenNPB81a,NielsenNPB81b} Hence the
edge modes must merge into the excitation spectrum at finite $q$, and the
gapless points occur either at $q=0$ or $q=\pi $, but not at both.\cite%
{MaoJPSJ10}

As the sample width is decreased, due to the finite size effect, the wave
functions of the edge modes on opposite edges overlap and mix, thus the
spectrum reopens a gap. In Sec. III, we will focus on this issue.

\section{Majorana edge states in the boundaries of $p$-wave superconductors}

Here we introduce a two-component field operator for each site $\alpha $
along the $y$-axis direction:%
\begin{equation}
\hat{d}_{\alpha }\left( q\right) =\frac{1}{2}\left[
\begin{array}{c}
\gamma _{2\alpha -1}\left( q\right) \\
\gamma _{2\alpha }\left( q\right)%
\end{array}%
\right] .
\end{equation}%
The Hamiltonian (\ref{H3}) can be expressed in the terms of two component
creation and annihilation operators $\hat{d}_{\alpha }^{\dag }\left(
q\right) $ and $\hat{d}_{\alpha }\left( q\right) $ as
\begin{equation}
H=\sum_{q,\alpha }\left\{ \hat{d}_{\alpha }^{\dag }\left( q\right) \mathcal{%
\hat{E}}\hat{d}_{\alpha }\left( q\right) +\left[ \hat{d}_{\alpha }^{\dag
}\left( q\right) \mathcal{\hat{T}}\hat{d}_{\alpha +1}\left( q\right) +h.c.%
\right] \right\} ,  \label{H4}
\end{equation}%
where%
\begin{equation}
\mathcal{\hat{E}}=-2\left\vert \Delta \right\vert \sin q\sigma _{z}-\left(
\mu +2t\cos q\right) \sigma _{y},
\end{equation}%
\begin{equation}
\mathcal{\hat{T}}=-t\sigma _{y}-i\left\vert \Delta \right\vert \sigma _{x},
\end{equation}%
with $\sigma _{i}$ ($i=x,y,z$) being the Pauli matrices. We define the
single particle states of the form%
\begin{equation}
\Psi \left( q\right) =\sum_{\alpha }\hat{d}_{\alpha }^{\dag }\left( q\right)
\psi _{\alpha }\left\vert 0\right\rangle ,
\end{equation}%
where $\left\vert 0\right\rangle $ is the vacuum state annihilated by the
operator $\hat{d}_{\alpha }\left( q\right) $, and $\psi _{\alpha }=\left(
\psi _{\alpha 1},\psi _{\alpha 2}\right) ^{T}$ is the two-component
amplitude with index $\alpha $. An open boundary condition $\psi _{\alpha
=0}=\psi _{\alpha =n+1}=0$ is introduced along the sample width direction.
The corresponding Schr\"{o}dinger equation is given by%
\begin{equation}
\mathcal{\hat{E}}\left( q\right) \psi _{\alpha }+\mathcal{\hat{T}}\left(
q\right) \psi _{\alpha +1}+\mathcal{\hat{T}}^{\dag }\left( q\right) \psi
_{\alpha -1}=\varepsilon \left( q\right) \psi _{\alpha }.  \label{SE1}
\end{equation}%
Note that Eq. (\ref{SE1}) is analogue to the equation given in studying edge
modes of topological insulators based on the tight-binding model.\cite%
{MaoJPSJ10,MaoPRB11,KonigJPSJ08}

We study solutions of the form $\psi _{\alpha }\sim \lambda ^{\alpha }\psi $
with $\lambda $ being a general complex number.\cite{CreutzPRD94} Firstly
substituting the ansatz into Eq. (\ref{SE1}), we have
\begin{equation}
\left[ \mathcal{\hat{E}}\left( q\right) +\lambda \mathcal{\hat{T}}\left(
q\right) +\lambda ^{-1}\mathcal{\hat{T}}^{\dag }\left( q\right) \right] \psi
=\varepsilon \left( q\right) \psi .  \label{SE2}
\end{equation}%
Thus the secular equation gives an equation about $\lambda $ and $%
\varepsilon \left( q\right) $
\begin{eqnarray}
&&\left( t^{2}-\left\vert \Delta \right\vert ^{2}\right) z^{2}+2t\left( \mu
+2t\cos q\right) z  \notag \\
&=&\varepsilon ^{2}-4\left\vert \Delta \right\vert ^{2}\left( 1+\sin
^{2}q\right) -\left( \mu +2t\cos q\right) ^{2},  \label{Z}
\end{eqnarray}%
where $z=\left( \lambda +\lambda ^{-1}\right) $. Note that, if $\lambda $ is
a solution, so is $\lambda ^{-1}$. Thus for every exponentially increasing
solution, there exists another one which exponentially decreases. \cite%
{CreutzPRD94} Four roots of $\lambda $ are given by%
\begin{equation}
\lambda _{1,2}=\frac{z_{\pm }+\sqrt{z_{\pm }^{2}-4}}{2},\text{ \ }\lambda
_{3,4}=\frac{z_{\pm }-\sqrt{z_{\pm }^{2}-4}}{2},  \label{LB}
\end{equation}%
where $z_{\pm }$ are two roots of $z$ in Eq. (\ref{Z}). We define two of
four roots of $\lambda $ as $\lambda _{\pm }$ which satisfy the relation $%
\left\vert \lambda _{\pm }\right\vert <1$, and then the other two roots of $%
\lambda $ are $\lambda _{\pm }^{-1}$ satisfying $\left\vert \lambda _{\pm
}^{-1}\right\vert >1$.

Now we consider the edge states. For convenience, a symmetric boundary
condition is used in the following, that is $\psi _{\tilde{\alpha}%
=-Y/2}=\psi _{\tilde{\alpha}=Y/2}=0$. Here, $Y=n+1$, and the new index $%
\tilde{\alpha}=\alpha -(n+1)/2$. Thus for the wave function $\psi _{\tilde{%
\alpha}}=\left( \psi _{\tilde{\alpha}1},\psi _{\tilde{\alpha}2}\right) ^{T}$
we have an analytical expression\cite{ZhouPRL08} $\ $
\begin{equation}
\psi _{\tilde{\alpha}1}=\tilde{c}_{+}f_{+}\left( q,\tilde{\alpha}\right) +%
\tilde{c}_{-}f_{-}\left( q,\tilde{\alpha}\right) ,
\end{equation}%
\begin{equation}
\psi _{\tilde{\alpha}2}=\tilde{d}_{+}f_{+}\left( q,\tilde{\alpha}\right) +%
\tilde{d}_{-}f_{-}\left( q,\tilde{\alpha}\right) ,
\end{equation}%
where%
\begin{equation}
f_{+}\left( q,\tilde{\alpha}\right) =\frac{\cosh \left( \tilde{\lambda}_{+}%
\tilde{\alpha}\right) }{\cosh \left( \tilde{\lambda}_{+}Y/2\right) }-\frac{%
\cosh \left( \tilde{\lambda}_{-}\tilde{\alpha}\right) }{\cosh \left( \tilde{%
\lambda}_{-}Y/2\right) },
\end{equation}%
\begin{equation}
f_{-}\left( q,\tilde{\alpha}\right) =\frac{\sinh \left( \tilde{\lambda}_{+}%
\tilde{\alpha}\right) }{\sinh \left( \tilde{\lambda}_{+}Y/2\right) }-\frac{%
\sinh \left( \tilde{\lambda}_{-}\tilde{\alpha}\right) }{\sinh \left( \tilde{%
\lambda}_{-}Y/2\right) },
\end{equation}%
with $\tilde{\lambda}_{\pm }=\ln \lambda _{\pm }$. The nontrivial solution
for the coefficients $\tilde{c}_{\pm }$ and $\tilde{d}_{\pm }$ in the wave
functions leads to a secular equation%
\begin{eqnarray}
&&\left( t^{2}-\left\vert \Delta \right\vert ^{2}\right) \left( \cosh ^{2}%
\tilde{\lambda}_{+}+\cosh ^{2}\tilde{\lambda}_{-}\right) +2\left\vert \Delta
\right\vert ^{2}  \notag \\
&=&2t^{2}\cosh \tilde{\lambda}_{+}\cosh \tilde{\lambda}_{-}-T\left\vert
\Delta \right\vert ^{2}\sinh \tilde{\lambda}_{+}\sinh \tilde{\lambda}_{-},
\label{SE3}
\end{eqnarray}%
with
\begin{equation}
T=\frac{\tanh \left( \tilde{\lambda}_{+}Y/2\right) }{\tanh \left( \tilde{%
\lambda}_{-}Y/2\right) }+\frac{\tanh \left( \tilde{\lambda}_{-}Y/2\right) }{%
\tanh \left( \tilde{\lambda}_{+}Y/2\right) }.
\end{equation}%
Equations (\ref{LB}) and (\ref{SE3}) give the energy dispersions and the
values of two characteristic quantities $\lambda _{\pm }=e^{\tilde{\lambda}%
_{\pm }}$.

When the sample width is sufficiently large, i.e., in the limit of $%
n\rightarrow \infty $, we can find two chiral edge modes with energy%
\begin{equation}
\varepsilon _{\pm }\left( q\right) =\pm 2\left\vert \Delta \right\vert \sin
q,  \label{EE}
\end{equation}%
and
\begin{equation}
\lambda _{\pm }=\frac{-\left( \mu +2t\cos q\right) \pm \sqrt{\left( \mu
+2t\cos q\right) ^{2}-4t^{2}+4\left\vert \Delta \right\vert ^{2}}}{2\left(
t+\left\vert \Delta \right\vert \right) }.
\end{equation}%
For nonzero $q$, the eigenstates for $\varepsilon _{\pm }$ $>0$ become
concentrated on one edge or the other, depending on the sign of $q$, that is
the left-moving and right-moving edge modes are localized at the opposite
boundaries, respectively.\cite{ReadPRB00} Equation (\ref{EE}) is consistent
with the result by numerical diagonalization (to see Fig. 1). Interesting to
note the cases of $q=0$ (for $\mu <0$ ) and $\pi $ (for $\mu >0$ ), one has
zero energy mode $\varepsilon \left( q=0/\pi \right) =0$ which corresponds
to the Majorana edge state in $p$-wave superconductors.\cite{ReadPRB00}

Actually, there are no exact $\varepsilon =0$ modes for a finite width
sample. One of the key features for the solution of a finite width is the
gap $E_{g}$ opening for the energy dispersion of the edge state. From Eqs. (%
\ref{LB}) and (\ref{SE3}), one has
\begin{eqnarray}
&&\varepsilon ^{2}-4\left\vert \Delta \right\vert ^{2}\sin ^{2}q  \notag \\
&=&2t\left( \mu +2t\cos q\right) \left( \cosh \lambda _{+}+\cosh \lambda
_{-}\right) +\left( \mu +2t\cos q\right) ^{2}  \notag \\
&&+4t^{2}\cosh \lambda _{+}\cosh \lambda _{-}-2T\left\vert \Delta
\right\vert ^{2}\sinh \lambda _{+}\sinh \lambda _{-},
\end{eqnarray}%
thus it can be found that a finite energy gap $E_{g}$ at $q=0$ (for the case
of $\mu <0$ ) is approximately
\begin{equation}
E_{g}\simeq \left\vert \Delta \right\vert \sqrt{\left\vert \frac{2\mu \left(
\mu +4t\right) }{t^{2}-\left\vert \Delta \right\vert ^{2}}\right\vert }%
e^{-n/l_{0}},
\end{equation}%
and at $q=\pi $ (for the case of $\mu >0$ ) \
\begin{equation}
E_{g}\simeq \left\vert \Delta \right\vert \sqrt{\left\vert \frac{2\mu \left(
\mu -4t\right) }{t^{2}-\left\vert \Delta \right\vert ^{2}}\right\vert }%
e^{-n/l_{0}},
\end{equation}%
where $l_{0}^{-1}=\min (\left\vert \ln \left\vert \lambda _{+}\right\vert
\right\vert ,\left\vert \ln \left\vert \lambda _{-}\right\vert \right\vert )$%
, and $l_{0}$ denotes the localization length of the edge mode. The gap $%
E_{g}$ will decrease exponentially with increase of the width of the strip.

\section{Majorana end states in quasi-1D $p$-wave superconductors}

In a recent work by Potter and Lee,\cite{PotterPRL10} two well-isolated
Majorana end states localized at opposite ends of the strip geometry have
been obtained by numerical diagonalization of the Hamiltonian (\ref{H1}).
Here we will give the phase diagrams of the presence of Majorana end modes
in quasi-1D $p$-wave superconductors by using topological arguments due to
Kitaev.\cite{KitaevPU01} To this aim, we consider the $2n\times 2n$ matrix $%
B\left( q\right) $ in the Hamiltonian (\ref{H3}). The matrix $B$ is an
antisymmetric matrix when $q$ is equal to zero or $\pi $, such that we can
calculate the Pfaffians Pf$B\left( 0\right) $ and Pf$B\left( \pi \right) $.
The topological property of the system described by the Hamiltonian (\ref{H3}%
) is characterized by a $Z_{2}$ topological index (Majorana number) $%
\mathcal{M}$:
\begin{equation}
\mathcal{M}=\text{sgn}\left[ \text{Pf}B\left( 0\right) \right] \text{sgn}%
\left[ \text{Pf}B\left( \pi \right) \right] =\pm 1,
\end{equation}%
where $+1$ corresponds to topologically trivial states and $-1$ to
topologically nontrivial states (i.e., the existence of zero mode Majorana
end states).\cite{KitaevPU01,LutchynPRL11,LutchynPRL10,GhoshPRB10}

For the simplest case, there is only one lattice site along the $y$-axis
direction (i.e., $n=1$). This case is just the Kitaev original model.\cite%
{KitaevPU01} Two $2\times 2$ antisymmetric matrices are
\begin{equation}
B_{n=1}\left( 0/\pi \right) =%
\begin{bmatrix}
0 & \mu \pm 2t \\
-\left( \mu \pm 2t\right) & 0%
\end{bmatrix}%
,
\end{equation}%
and Pf$B_{n=1}\left( 0/\pi \right) =\mu \pm 2t$, where "$+$" and "$-$"
correspond to the cases of $q=0$ and $\pi $, respectively. The Majorana
number for the case of the strict 1D limit is given
\begin{equation}
\mathcal{M}_{n=1}=\text{sgn}\left( \mu +2t\right) \text{sgn}\left( \mu
-2t\right) ,
\end{equation}%
thus we have the topologically nontrivial condition%
\begin{equation}
2\left\vert t\right\vert >\left\vert \mu \right\vert \text{ }\left(
\left\vert \Delta \right\vert \neq 0\right) .  \label{n1}
\end{equation}%
The above Eq. (\ref{n1}) is just the result by Kitaev,\cite{KitaevPU01} who
demonstrated for a long open chain (in the limit of $L\rightarrow \infty $)
there are zero energy Majorana end states localized near per boundary point
under the condition (\ref{n1}). However, if the chain length $L$ is finite,
there is a weak interaction between two unpaired Majorana fermions.\cite%
{KitaevPU01} Following the method given in Sec. III, we can also obtain the
energy gap opened due to the finite size effect approximately as
\begin{equation}
\Delta _{g}\simeq \left\vert \Delta \right\vert \sqrt{\left\vert \frac{%
2\left( 4t^{2}-\mu ^{2}\right) }{t^{2}-\left\vert \Delta \right\vert ^{2}}%
\right\vert }e^{-L/\tilde{l}_{0}},
\end{equation}%
where $\tilde{l}_{0}^{-1}=\min (\left\vert \ln \left\vert x_{+}\right\vert
\right\vert ,\left\vert \ln \left\vert x_{-}\right\vert \right\vert )$, with%
\begin{equation}
x_{\pm }=\frac{-\mu \pm \sqrt{\mu ^{2}-4t^{2}+4\left\vert \Delta \right\vert
^{2}}}{2\left( t+\left\vert \Delta \right\vert \right) }.
\end{equation}%
Here, $\tilde{l}_{0}$ indicates the localization length of the Majorana end
states. Thus, the energy gap $\Delta _{g}$ vanishes as $\exp \left( -L/%
\tilde{l}_{0}\right) $ in an open chain.\cite{KitaevPU01}

For the case of $n=2$, the lattice site numbers along the $y$-axis direction
are two. Two $4\times 4$ antisymmetric matrices are
\begin{eqnarray}
&&B_{n=2}\left( 0/\pi \right)  \notag \\
&=&%
\begin{bmatrix}
0 & \mu \pm 2t & 0 & t-\left\vert \Delta \right\vert \\
-\left( \mu \pm 2t\right) & 0 & -\left( t+\left\vert \Delta \right\vert
\right) & 0 \\
0 & t+\left\vert \Delta \right\vert & 0 & \mu \pm 2t \\
-\left( t-\left\vert \Delta \right\vert \right) & 0 & -\left( \mu \pm
2t\right) & 0%
\end{bmatrix}%
.
\end{eqnarray}%
The direct calculation yields the Pfaffians Pf$B_{n=2}\left( 0/\pi \right) $
\begin{equation}
\text{Pf}B_{n=2}\left( 0/\pi \right) =\left( \mu \pm 2t\right)
^{2}+\left\vert \Delta \right\vert ^{2}-t^{2}.
\end{equation}

For the lager lattice site numbers $n$ ($\geq 3$), Pf$B_{n}\left( 0/\pi
\right) $ can be also calculated analytically, and we obtain a recursion
relation
\begin{equation}
\text{Pf}B_{n}\left( 0/\pi \right) =a_{\pm }\text{Pf}B_{n-1}\left( 0/\pi
\right) +b\text{Pf}B_{n-2}\left( 0/\pi \right) ,  \label{Pf}
\end{equation}%
where $a_{\pm }=\mu \pm 2t$ and $b=\left\vert \Delta \right\vert ^{2}-t^{2}$%
. We further solve Eq. (\ref{Pf}), and give an analytic formula for Pf$%
B_{n}\left( 0/\pi \right) $%
\begin{equation}
\text{Pf}B_{n}\left( 0/\pi \right) =\frac{\left(
r_{1}^{n+1}-r_{2}^{n+1}\right) }{\sqrt{a_{\pm }^{2}+4b}},
\end{equation}%
where%
\begin{equation}
r_{1}=\frac{a_{\pm }+\sqrt{a_{\pm }^{2}+4b}}{2},\text{ \ }r_{2}=\frac{a_{\pm
}-\sqrt{a_{\pm }^{2}+4b}}{2}.
\end{equation}%
According to the Pfaffians Pf$B_{n}\left( 0/\pi \right) $ one can compute $%
\mathcal{M}$ as a function of the physical parameters, and then plot the
phase diagram showing a sequence of topological phase transition for
different the lattice site numbers $n$. Figures 2 and 3 plot the phase
diagrams for the even and odd lattice site numbers $n$ along the $y$-axis
direction, respectively. The phase diagrams of this tight-binding model have
the symmetry on positive and negative $\mu $ values, thus here we only plot
on negative $\mu $ values because the other part on positive $\mu $ values
is a mirror image. However, this $\mu \rightarrow -\mu $ symmetry is not
generic to models with say, next-nearest-neighbor hopping or
next-nearest-neighbor pairing.

\begin{figure}[tbp]
\includegraphics[width=8cm]{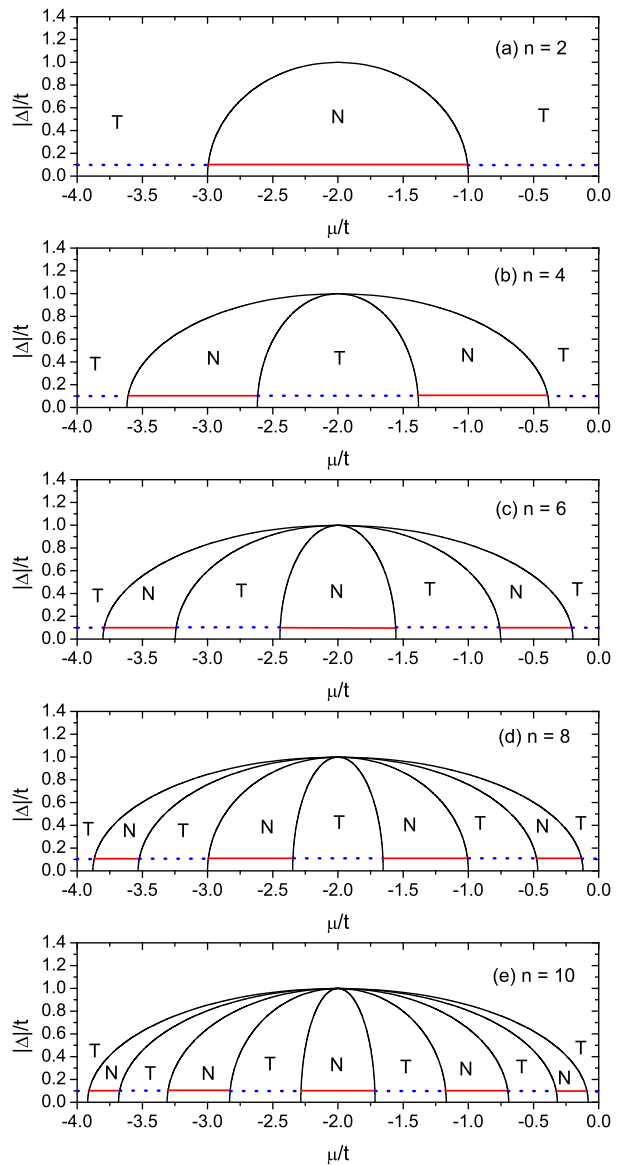}
\caption{{}(color online). Phase diagram for the quasi-1D $p$-wave
superconductor model as a function of the $p$-wave pairing amplitude and
chemical potential for even lattice site numbers $n$ along the $y$-axis
direction. "N" denotes the topologically nontrivial region in the present of
zero mode Majorana end states and "T" denotes the topologically trivial
region without zero mode states. When $\left\vert \Delta \right\vert /t=0.1$%
, the solid (red) lines and dotted (blue) lines guide the values of $\protect%
\mu /t$ corresponding to the topologically nontrivial and trivial phases,
respectively. \ (a) $n=2$; (b) $n=4$; (c) $n=6$; (d) $n=8$; (e) $n=10$.}
\end{figure}

\begin{figure}[tbp]
\includegraphics[width=8cm]{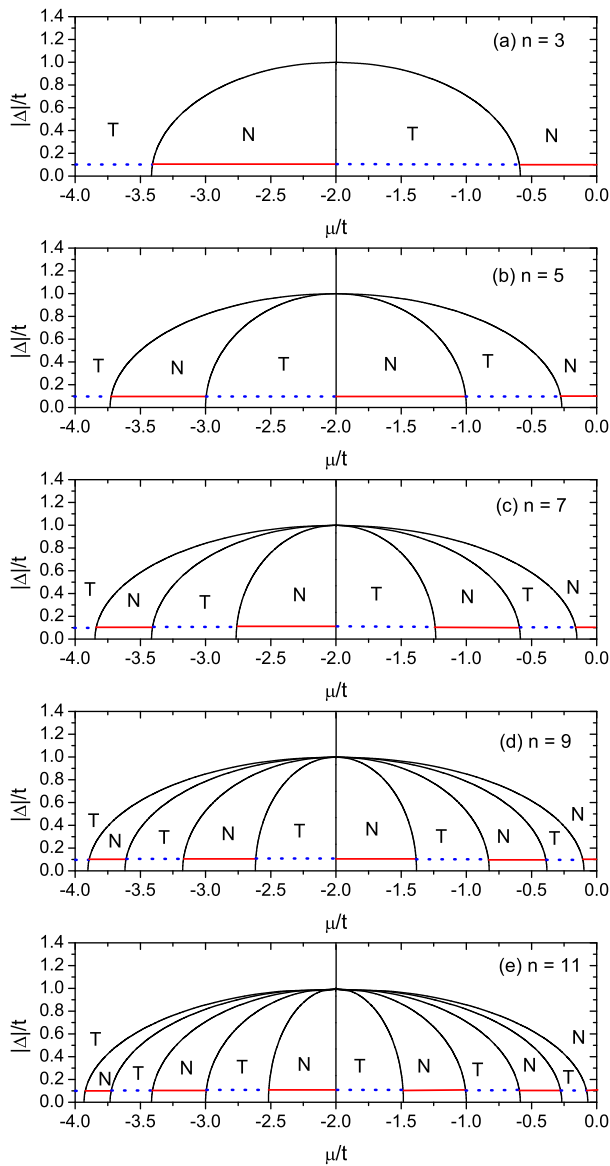}
\caption{{}(color online). Phase diagram for the quasi-1D $p$-wave
superconductor model as a function of the $p$-wave pairing amplitude and
chemical potential for odd lattice site numbers $n$ along the $y$-axis
direction. "N" denotes the topologically nontrivial region in the present of
zero mode Majorana end states and "T" denotes the topologically trivial
region without zero mode states. When $\left\vert \Delta \right\vert /t=0.1$%
, the solid (red) lines and dotted (blue) lines guide the values of $\protect%
\mu /t$ corresponding to the topologically nontrivial and trivial phases,
respectively. \ (a) $n=3$; (b) $n=5$; (c) $n=7$; (d) $n=9$; (e) $n=11$.}
\end{figure}

We now analyze these phase diagrams. Firstly, it is observed that for the
given value of $\left\vert \Delta \right\vert /t$ the topologically
nontrivial and trivial phases alternate with the variation of the value of $%
\mu /t$. As an example, the case of $\left\vert \Delta \right\vert /t=0.1$
is shown in the Figs. 2 and 3. From the Figs. 2 and 3, it is shown that the
phase diagrams have different properties for the even and odd lattice site
numbers $n$. Such the odd-even effect should be an artifact of the
tight-binding model and may not be generic to all quasi-1D p-wave
supercondurtors, since the order parameter for Cooper pairing are presented
as a model parameter, not given by a self-consistent mean field calculation.
Even in a realistic material, the oscillation of the critical temperature
with the monolayer number of the sample was observed experimentally as
quantum confinement effect.\cite{GuoScience04} It is not clear what the real
meaning of odd and even $n$ in this model calculation, which may have no
counterpart in the continuous model as a long wave length limit. However,
the oscillation of the presence and absence of the zero mode still clearly
exhibits in the continuous model as quantum confinement effect.
\begin{figure}[tbp]
\includegraphics[width=8cm]{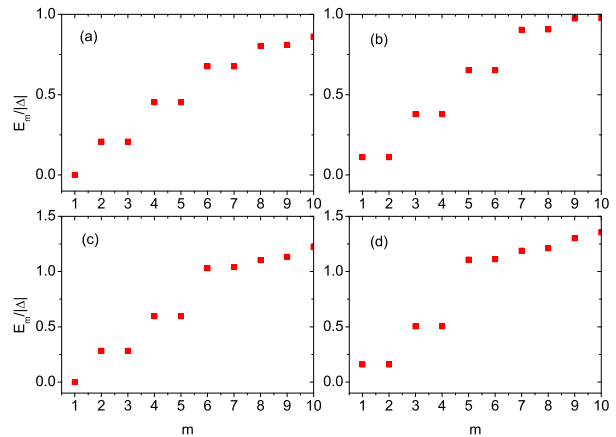}
\caption{{}(color online). Excitation spectrum $E_{m}/\left\vert \Delta
\right\vert $ for a square-lattice Kitaev model obtained by numerical
diagonalization of the Hamiltonian (\protect\ref{H1}) with parameters: $L=100
$, $n=10$, $\left\vert \Delta \right\vert /t=0.1$. $m$ labels eigenvalues of
the Hamiltonian (\protect\ref{H1}). Here only the partial low excitation
energies are presented. (a) $\protect\mu /t=-1.0$ ($E_{1}/\left\vert \Delta
\right\vert =1.09\times 10^{-5}$); (b) $\protect\mu /t=-1.5$ ($%
E_{1}/\left\vert \Delta \right\vert =0.1116$); (c) $\protect\mu /t=-2.0$ ($%
E_{1}/\left\vert \Delta \right\vert =1.54\times 10^{-5}$); (e) $\protect\mu %
/t=-2.5$ ($E_{1}/\left\vert \Delta \right\vert =0.1595$).}
\end{figure}

\begin{figure}[tbp]
\includegraphics[width=8cm]{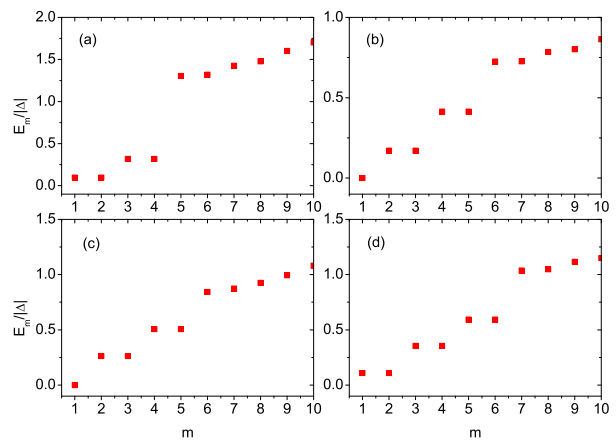}
\caption{{}{}(color online). Excitation spectrum $E_{m}/\left\vert \Delta
\right\vert $ for a square-lattice Kitaev model obtained by numerical
diagonalization of the Hamiltonian (\protect\ref{H1}) with parameters: $%
L=100 $, $\left\vert \Delta \right\vert /t=0.1$, $\protect\mu /t=-1.1$. $m$
labels eigenvalues of the Hamiltonian (\protect\ref{H1}). Here only the
partial low excitation energies are presented. (a) $n=6$ ($E_{1}/\left\vert
\Delta \right\vert =$ $0.0966$); (b) $n=7$ ($E_{1}/\left\vert \Delta
\right\vert =8.21\times 10^{-6}$); (c) $n=8$ ($E_{1}/\left\vert \Delta
\right\vert =7.39\times 10^{-6}$); (e) $n=9$ ($E_{1}/\left\vert \Delta
\right\vert =0.1102$).}
\end{figure}

According to the topological argument above, the topologically nontrivial
phase corresponds to the existence of the zero energy mode, which can be
justified by the excitation spectrum obtained by the numerical
diagonalization of the Hamiltonian (\ref{H1}). According to the work by
Potter and Lee,\cite{PotterPRL10} for sufficiently long samples ($L\gg
e^{n/\xi _{0}}$), one can find spatially isolated zero energy Majorana end
states localized at opposite ends of the sample with the strip geometry. To
see Fig. 2(e) (with $n=10$), we can find that if $\left\vert \Delta
\right\vert /t$ is fixed at $0.1$, the cases of $\mu /t=-1.0$ and $-2.0$
correspond to the topologically nontrivial phase; the cases of $\mu /t=-1.5$
and $-2.5$ to the topologically trivial phase. For the above four parameter
cases, the excitation spectra of the Hamiltonian (\ref{H1}) with $L=100$ and
$n=10$ are plotted in Fig. 4.\ Numerical diagonalizations show that the zero
energy modes indeed exist in the cases of $\mu /t=-1.0$ and $-2.0$; there
are no zero energy mode in the cases of $\mu /t=-1.5$ and $-2.5$. The
conclusions are consistent with the results by Potter and Lee.\cite%
{PotterPRL10}

We note also an interesting property from the phase diagrams. For some given
parameters, the topological property of a strip geometry will change in an
oscillatory way with respect of the sample width. For instance, in Fig. 2,
when the parameters $\left\vert \Delta \right\vert /t=0.1$ and $\mu /t=-2.0$
are fixed, the topologically nontrivial and trivial phases alternate when
the lattice site numbers $n$ (even numbers) along the $y$-axis direction
change from $n=2$ to $n=10$. While in the cases of the odd $n$, $\mu /t=-2.0$
is a topological phase transition line (to see Fig. 3.) As an other example,
we take the parameters $\left\vert \Delta \right\vert /t=0.1$, $\mu /t=-1.1$%
, and then from the Figs. 2 and 3 it is presented that the topologically
nontrivial phases exist in the cases of $n=2$, $4$, $5$, $7$, $8$, $10$, and
$11$; while the cases of $n=3$, $6$, and $9$ correspond to the topologically
trivial phases. In Fig. 5 the excitation spectra for four different sample
widths ($n=6$, $7$, $8$, and $9$) are shown. For the case $n=7$ and $8$,
there are zero energy modes in the given parameters; while the zero energy
modes disappear in the case $n=6$ and $9$. Thus the topological property
obtained by the phase diagrams are consistent with the results by numerical
diagonalization.

\section{Conclusions}

In this paper, we investigate crossover from Majorana edge to end states in
an ideal spinless quasi-1D $p$-wave superconductor based on the
tight-banding model. We found the existence of Majorana edge modes when the
sample width is the sufficiently large, and then through analytical
solutions in a model calculation two Majorana fermions at the two edges can
couple together to produce a gap in the excitation spectrum for a strip of
finite width under the periodic boundary condition along the longitudinal
direction. We calculate $Z_{2}$ topological index in the quasi-1D sample,
and plot the phase diagrams of the presence of Majorana end modes. By
analyzing the phase diagrams, we find that for some given parameters the
topological property of a strip geometry will change in an oscillatory way
with respect of the sample width.

\begin{acknowledgments}
This work was supported by the Research Grant Council of Hong Kong under
Grant Nos. HKU7051/10P, and HKUST3/CRF/09. ZB was supported by was supported
by National Natural Science Foundation of China (Grant No. 10974046), Hubei
Provincial Natural Science Foundation of China (Grant No. 2009CDB360), and
the Key Project of Education Department of Hubei Province of China (Grant
No. D20101004).
\end{acknowledgments}

\end{document}